\documentclass[
preprint,
showpacs,                     
showkeys,                  
secnumarabic,
amssymb, 
nobibnotes, 
aps, 
pre]{revtex4-1}

\usepackage{graphicx}
\usepackage{latexsym}
\usepackage{amsmath}
\usepackage{verbatim}

\begin{document}

\title{The thermodynamic entropy 
of a macroscopic quantum system 
is a continuous function of energy}

\author{Robert H. Swendsen}
\email[]{swendsen@cmu.edu}
\affiliation{Department of Physics, Carnegie Mellon University, Pittsburgh PA, 15213, USA}

                              
\date{\today}


\begin{abstract}
The proper definition of entropy 
is fundamental to the relationship between
statistical mechanics and thermodynamics.
It also 
plays a major role in the recent debate about 
the validity of the concept of negative temperature.
In this paper,
I analyze and calculate 
the thermodynamic entropy 
for large, but finite
quantum  mechanical systems.
A special feature  of this analysis 
is that the   thermodynamic energy  of a quantum system 
is shown to be  a continuous variable,
rather than being associated with  discrete energy eigenvalues.
Calculations of the entropy as a function of energy
can be carried out  with a Legendre transform 
of thermodynamic potentials obtained from a canonical ensemble.
The resultant expressions for the entropy 
are also able to describe
equilibrium between
 quantum systems 
having incommensurate 
 energy-level spacings.
 This definition of entropy preserves all required thermodynamic properties,
including satisfaction of  all
postulates 
and   laws of thermodynamics.
It also demonstrates the consistency of the concept 
of negative temperature with the principles of thermodynamics.
\end{abstract}

\maketitle

\section{Introduction}

In recent years,
there has been a  renewal 
of interest in the fundamental definition of
the thermodynamic entropy
in statistical mechanics.
An important source of that interest has been the 
controversial claims that the 
Gibbs ``volume'' entropy
(defined below)
is the only valid definition\cite{DH_Physica_A_2006,Vilar_Rubi_2014,Sokolov_2014,HHD_2014,Campisi_SHPMP_2015,HHD_2015},
and that the concept of 
negative temperature
is inconsistent with thermodynamics\cite{Purcell_Pound,Ramsey_Neg_T,Landsberg_1959,DH_2013,Romero-Rochin_2013,DH_NatPhys_2014,Schneider_et_al,Frenkel_Warren_2015,DH_reply_to_FW}. 
Earlier work on classical systems
has shown 
 how a valid definition of entropy 
 is consistent with negative temperatures,
 giving correct thermodynamic predictions 
when the volume entropy does not\cite{RHS_1,RHS_4,RHS_5,RHS_8,RHS_9,RHS_unnormalized,SW_2015_AJP,SW_2015_PR_E_R}.

The entropy of quantum systems is complicated by a discrete energy spectrum,
which has led many to regard the entropy as being defined only at energies corresponding to 
eigenstates\cite{Planck_1901,Planck_book}, 
or as constants between eigenvalue energies (step functions)\cite{HHD_2014,Campisi_SHPMP_2015,HHD_2015}.
 In this paper, 
 I argue  that the  thermodynamic energy
 of  macroscopic quantum systems
has 
a continuous spectrum of  values,
even though the eigenvalues of the energy form a discrete set.
To briefly summarize my arguments,
macroscopic measurements 
never put a many-body system into an energy eigenstate.
The small interactions
that are necessary to establish thermodynamic equilibrium
 between two or more systems
are sufficient to alter the discrete energy spectra of
the isolated systems.
This means that  the projection 
of an eigenstate of the  composite system of interacting subsystems
onto the energy levels of an isolated subsystem
will never produce an eigenstate
of the isolated subsystem.
An isolated system will necessarily be described 
by an ensemble of quantum states,
including contributions from many energy eigenvalues.

I  begin by demonstrating 
in Section~\ref{section: QM}
that the thermodynamic energy 
of an isolated quantum system that was previously
in thermal contact with another system
is a continuous variable.
Section~\ref{section: canonical}
 reviews the canonical probability distribution of energy 
in a system that has been in contact with a thermal reservoir,
that is, with a very large (formally, infinite) system.
I  analyze the canonical distribution 
in terms of a Massieu function\cite{Callen,RHS_book},
in order to be able to carry out calculations for 
either a monotonic or a non-monotonic 
energy density of states.
The essential features of an ensemble for an isolated system
are then  shown to depend on an internal temperature variable,
which leads to a method for calculating the entropy 
of a  macroscopic quantum  system.
As examples,
explicit expressions 
are derived
for the entropy
of 
 quantum simple harmonic oscillators
and  two-level objects.
The entropy of a system of two-level objects 
has a non-monotonic density of states,
and serves to demonstrate that 
negative temperatures
are consistent with the principles 
of both 
 quantum statistical mechanics
and
thermodynamics, as previously shown
for the classical case 
 in Ref.~\onlinecite{SW_2015_PR_E_R}.

\section{Thermodynamic energy in quantum systems}
\label{section: QM}

To see the unique way in which  interactions
affect thermal ensembles 
in quantum statistical mechanics
and lead to the thermodynamic energy 
being a continuous function,
 consider  a composite system 
 with $M$ 
 macroscopic  (many-body) subsystems.
 The macroscopic observables 
 for the composite system are just 
 the set of $3M$ variables
$\{$\textbf{E}$, $\textbf{V}$,$\textbf{N}$\}=\{ E_j, V_j,N_j \vert j= 1, \cdots,M \}$,
where $\textbf{E}$,$\textbf{V}$, and $\textbf{N}$
denote the sets of energy, volume, and particle numbers.
To reduce the proliferation of subscripts,
the equations  are written for a single type of particle,
but the generalization to many types of particles 
is obvious.
For quantum spins,
the magnetization will also be included in the 
macroscopic variables.

Each macroscopic subsystem $j$ will be characterized by 
a Hamiltonian 
$H_j$,
\emph{when it is isolated from other subsystems}.
To establish thermal equilibrium 
between subsystems,
direct
interactions 
between particles in different subsystems
are necessary.
These additional interactions 
will be denoted by the interaction Hamiltonians
$\textbf{L}=\{ L_{j,k} \vert j= 1, \cdots,M-1;  k= j, \cdots,M; \}$.
These interactions 
between macroscopic subsystems
prove to be essential  
in  deriving  the quantum entropy.

The extension of the definition of entropy 
to quantum systems follows the  basic pattern 
of considering thermodynamic interactions 
between subsystems of a composite system
that I discussed  for classical composite systems in earlier work\cite{RHS_unnormalized}.
However,
 there are some subtleties 
associated with the discreteness of the 
spectrum of energy eigenvalues 
for finite systems.
 In particular, 
 the effects of
 interactions between subsystems,
 denoted by 
 $\textbf{L}$,
 must be treated carefully.
I  begin by introducing 
a fairly standard notation 
for
 quantum states and
 equilibrium quantum ensembles,
 first for isolated subsystems,
 and then for interacting subsystems.

\subsection{Isolated  quantum subsystems}
\label{subsection: QM notation}

First consider the case in which 
 $\textbf{L}=0$,
so that
every  macroscopic subsystem is isolated.
Denote the quantum Hamiltonian 
for the $j$-th subsystem
by $H_j$,
and the corresponding quantum numbers by
$\nu_j$.
The energy eigenvalue equation is then
\begin{equation}
H_j \vert \nu_j \rangle 
=
\epsilon_j  (\nu_j ) \vert \nu_j \rangle  ,
\end{equation}
where 
$\vert \nu_j \rangle $
denotes an eigenstate.
I assume that 
for each subsystem $j$,
the 
set of 
$\vert \nu_j \rangle $
form a complete, orthonormal set.

A general quantum state of subsystem  $j$
will be denoted by 
$\vert \psi_j \rangle$.
It can be expanded in the eigenstates of the subsystem,
\begin{equation}\label{expand psi}
\vert \psi_j \rangle
=
\sum_{\nu_j} c_{\nu_j} \vert \nu_j \rangle  ,
\end{equation}
where the $c_{\nu_j}$'s
are complex numbers,
$c_{\nu_j}= \left\vert c_{\nu_j} \right\vert  \exp \left( i \phi_{\nu_j} \right)$.
\begin{equation}
c_{\nu_j }
=
\langle \nu_j \vert \psi_j \rangle
\end{equation}
If 
$\vert \psi_j \rangle$ is properly normalized,
we have 
\begin{equation}
\sum_{\nu_j} \vert c_{\nu_j}   \vert^2
=
1  .
\end{equation}

Assuming  that there is a
probability distribution, 
$P_j( \psi_j )$,
defined on the set of quantum states,
then the average of an arbitrary operator $A$
will be given by 
\begin{equation}
\langle \langle A \rangle \rangle
=
\sum_{\psi_j} P_j(\psi_j) \langle  \psi_j \vert A \vert \psi_j   \rangle   ,
\end{equation}
where the double angular brackets indicate
that  both 
a quantum expectation value and 
an ensemble average are being calculated,
and the summation symbol is intended to cover 
whatever sums and integrals are appropriate.
Using 
Eq.~(\ref{expand psi}),
$\langle \langle A \rangle \rangle$ 
 can be written as 
\begin{equation}
\langle \langle A \rangle \rangle
=
\sum_{\{c_j\}} 
\sum_{\mu_j}
\sum_{\nu_j}
P_j( {\{c_j\}} )  c^*_{\mu_j} c_{\nu_j}   \langle  \mu_j \vert A \vert \nu_j \rangle   . 
\end{equation}
where 
$\{c_j\}$ 
indicates a set of values of 
$c_{\nu_j}$
corresponding to the quantum state 
$\vert \psi_j \rangle $,
and the sum over 
$\{c_j\}$ 
is the sum over all quantum states in the ensemble.

In equilibrium,
 all phases, $\phi_j$, are equally probable.
Integrating over the phases gives a factor of zero for all cross terms,
leaving  the expression
\begin{equation}
\langle \langle A \rangle \rangle
=
\sum_{\nu_j}
\sum_{\{c_j\}} 
P_j( {\{c\}} )  \vert c_{\nu_j} \vert^2   \langle  \nu_j \vert A \vert \nu_j \rangle   . 
\end{equation}
This expression can be simplified by writing
\begin{equation}
P_j(\nu_j ) = \sum_{\{c_j\}}   P_j( {\{c_j\}} ) \, \vert c_{\nu_j} \vert^2  ,
\end{equation}
so that 
$P(\nu_j )  \ge 0$.
This gives 
\begin{equation}
\langle \langle A \rangle \rangle
=
\sum_{\nu_j}
P_j(\nu_j ) \, \langle  \nu_j \vert A \vert \nu_j \rangle   . 
\end{equation}
Further noting that 
$\langle \langle 1 \rangle \rangle = 1$,
we see that 
\begin{equation}
\sum_{\nu_j}
P_j(\nu_j )   =  1     ,
\end{equation}
so that 
$P_j(\nu_j ) $
acts very much like a probability.
Averages 
over the equilibrium quantum ensemble 
can be computed 
as if 
$P_j(\nu_j ) $
were a probability.
It is quite common to speak loosely of 
$P_j(\nu_j ) $
as being the probability of the subsystem being in the 
$\vert \nu_j \rangle$ 
eigenstate,
but 
it is well known 
this is not correct.
$P_j(\nu_j ) $
 is the probability that a precise measurement of the energy
would put the subsystem 
into the 
$\vert \nu_j \rangle$ 
eigenstate.
Since a macroscopic subsystem is generally in a state of the form given in 
Eq.~(\ref{expand psi}),
and
 macroscopic measurements 
cannot 
put a subsystem in an eigenstate,  
the  true  probability of a real macroscopic subsystem 
being in an energy eigenstate is zero.
This leads to the question 
of what we do know about the distribution of quantum states
in interacting subsystems.

\subsection{Interacting quantum  subsystems}
\label{subsection: QM interacting}

An important feature of interacting quantum subsystems 
is that the spectrum of energy eigenvalues 
is not  related to the corresponding spectra 
for the isolated subsystems in a simple way.
The composite  subsystem has new eigenfunctions
with new eigenvalues that depend 
on the interactions
$\textbf{L} \ne 0$.
As an example,
consider two interacting subsystems
with the following Hamiltonian.
\begin{equation}
H_{1,2} = H_1 +H_2 +L_{1,2}
\end{equation}
The eigenvalue equation for  this Hamiltonian can be written as 
\begin{equation}
H_{1,2} \vert \nu_{1,2} \rangle 
=
\epsilon_{1,2}  (  \nu_{1,2}  ) \vert \nu_{1,2} \rangle  ,
\end{equation}
where 
$\vert \nu_{1,2} \rangle $
denotes the eigenstate,
and 
$\nu_{1,2}$
the quantum number.
The eigenstates 
$\vert \nu_{1,2} \rangle $
form a complete, orthonormal set.

A general quantum state of the  composite system 
of interacting subsystems
will be denoted by 
$\vert \Psi_{1,2} \rangle$.
It can be expanded in the eigenstates of the composite  system.
\begin{equation}\label{expand Psi}
\vert \Psi_{1,2}  \rangle
=
\sum_{\nu_{1,2}  } C_{\nu_{1,2} } \vert \nu_{1,2}  \rangle  ,
\end{equation}
where the $C_{\nu_{1,2} }$'s
are complex numbers.
\begin{equation}
C_{\nu_{1,2} }
=
\langle \nu_{1,2} \vert \Psi_{1,2} \rangle
\end{equation}

If 
$\vert \Psi_{1,2}  \rangle$ is properly normalized,
we have 
\begin{equation}
\sum_{\nu_{1,2} }  \vert C_{\nu_{1,2} }   \vert^2
=
1     .
\end{equation}

\subsection{Energy distribution of an individual subsystem due to interactions}
\label{subsection: QM Ej}

Given 
the eigenstates and eigenvalues of the  composite system 
of interacting subsystems
and
the probability distribution
$P_{1,2} (\nu_{1,2})$,
we can compute the energy distribution 
in the individual subsystems.
If the  composite system is in a state
$\vert \Psi_{1,2}  \rangle$,
the projection onto the eigenstate 
$\vert \nu_1 \rangle$
 of the isolated subsystem $1$ is
\begin{equation}\label{nu PsI 1}
\langle \nu_1 \vert \Psi_{1,2}  \rangle
=
\sum_{\nu_{1,2}  } C_{\nu_{1,2} }   \langle \nu_1  \vert \nu_{1,2}  \rangle
\end{equation}
The expectation of the energy $E_1$ of subsystem 1,
given that the composite system is in state
$\vert \Psi_{1,2} \rangle$, 
is
\begin{eqnarray}\label{E1 ave 1}
\langle E_1 \rangle_{\Psi_{1,2}}
&=&
 \langle \Psi_{1,2}  \vert     H_1  \vert \Psi_{1,2}  \rangle  \nonumber \\
&=&
 \sum_{\mu_1}
 \sum_{\nu_1}
 \langle \Psi_{1,2}  \vert   \mu_1 \rangle
 \langle \mu_1 \vert
    H_1  
    \vert   \nu_1 \rangle
 \langle \nu_1  \vert \Psi_{1,2}  \rangle
\end{eqnarray}
Because 
$\vert \nu_1 \rangle$
is an eigenfunction of 
$H_1$,
Eq.~(\ref{E1 ave 1})
becomes
\begin{equation}\label{E1 ave 2}
\langle E_1 \rangle_{\Psi_{1,2}}
=
 \sum_{\nu_1}
 \langle \Psi_{1,2}  \vert   \nu_1 \rangle
 \langle \nu_1 \vert
    H_1  
    \vert   \nu_1 \rangle
 \langle \nu_1  \vert \Psi_{1,2}  \rangle
\end{equation}
or, using 
Eq.~(\ref{nu PsI 1}), 
\begin{equation}\label{E1 ave 3}
\langle E_1 \rangle_{\Psi_{1,2}}
=
 \sum_{\nu_1}
 \left\vert  C_{\nu_{1,2} }  \right\vert^2
 \left\vert \langle \nu_1  \right\vert \nu_{1,2}  \rangle  \vert^2
    E_1 (  \nu_1 )
\end{equation}

Averaging over the probability distribution of the 
states of the composite system
gives
\begin{equation}\label{E1 ave 4}
\langle\langle E_1 \rangle\rangle
=
 \sum_{\nu_1}
    E_1 (  \nu_1 )
\sum_{\nu_{1,2}}
 P_{1,2} ( \nu_{1,2} )
 \left\vert  C_{\nu_{1,2} }  \right\vert^2
 \left\vert \langle \nu_1  \right\vert \nu_{1,2}  \rangle  \vert^2
\end{equation}
This can be simplified by defining
\begin{equation}\label{P n1 n12 1}
p_1( \nu_1)
=
\sum_{\nu_{1,2}}
 P_{1,2} ( \nu_{1,2} )
 \left\vert  C_{\nu_{1,2} }  \right\vert^2
 \left\vert \langle \nu_1  \right\vert \nu_{1,2}  \rangle  \vert^2  .
\end{equation}
where 
$1 \ge p_1( \nu_1) \ge 0$,
and 
$\sum_{\nu_1} p_1( \nu_1) = 1$.
This gives 
\begin{equation}\label{E1 ave 5}
\langle\langle E_1 \rangle\rangle
=
 \sum_{\nu_1}
    E_1 (  \nu_1 )
p_1( \nu_1)   .
\end{equation}
The  value
$\langle\langle E_1 \rangle\rangle$,
including both quantum expectation values and 
ensemble averages,
corresponds to the thermodynamic energy,
$U_1$,
so that
 we can also write
 \begin{equation}\label{E1 ave 6}
U_1 
=
 \sum_{\nu_1}
    E_1 (  \nu_1 )
p_1( \nu_1) .
\end{equation}

The most important feature of 
Eqs.~(\ref{E1 ave 5}) and (\ref{E1 ave 6})
is that they demonstrate that 
 subsystems 
 that interact within a composite system are not in energy eigenstates.
This, in turn,
implies that the 
thermodynamic energy 
of a quantum subsystem 
in equilibrium with another quantum subsystem
can vary continuously,
even though the energy spectra are  discrete.
$U_1$
remains a continuous variable
when subsystems in thermal equilibrium 
are separated.

It is appropriate to compare the magnitudes of 
typical energy spacings with typical thermal fluctuations
of the energy.
In an example presented below,
we will consider a subsystem composed 
of $N$ two-level quantum objects,
with a level spacing of 
$\epsilon$.
Characteristic average energies are of the order of 
$N \epsilon$,
so that the relative size of the level spacing 
in comparison with the total energy
is of the order of 
$1/N$.
In contrast,
relative fluctuations of the energy 
are typically of the order of 
$1/\sqrt{N}$.
For a macroscopic subsystem with 
$N \approx 10^{20}$,
the thermal fluctuations are roughly
$10^{10}$ larger than the energy level spacing.

It is also important that typical energies 
associated with the interactions
$\textbf{L}$ are expected to scale with the surface area 
of the subsystem,
so that their relative magnitude goes as 
$N^{-1/3}$.
This is can easily be larger than the thermal fluctuations,
and much larger than the energy-level spacing.
As a result,
there is nothing to prohibit 
quantum subsystems with incommensurate 
energy-level spacings 
from exchanging energy and being in 
thermal equilibrium with each other.
Indeed, the explicit forms for the entropies of model subsystems
that are derived 
in Section~\ref{section: examples}
 can be used to 
predict thermal behavior due to interactions
between
macroscopic subsystems
with incommensurate energy-level spacings.

Another consequence of 
Eq.~(\ref{E1 ave 6})
is that the concept of 
a microcanonical ensemble 
must be modified 
in quantum statistical mechanics,
as discussed
in the next subsection.

\subsection{Separated quantum subsystems}

When two or more interacting 
\emph{classical} subsystems are separated and isolated,
they will each go into a microscopic state
with a unique energy.
Naturally,
the value of that energy can only be predicted
to within the thermal fluctuations of the energy 
before separation.
Nevertheless,
it  clearly makes sense to describe an isolated classical subsystem 
by a
microcanonical ensemble
with an exact value for the energy,
regardless of its history.

The situation when
\emph{quantum} subsystems are separated
is quite different.
Each subsystem will  go into a quantum state
that can be expressed as a linear combination of 
eigenstates,
as shown in 
Eq.~(\ref{expand psi}).
However, 
Eq.~(\ref{E1 ave 6})
 shows
this quantum state  will not be in an energy eigenstate,
so the subsystems cannot be described by the usual
microcanonical ensemble
with an exact value for the energy.
As demonstrated in the next section,
it is still possible to compute the fundamental relation,
$S_j=S_j(U_j,V_j,N_j)$,
for each subsystem
by using the canonical ensemble,
which is  valid for quantum statistical mechanics.
Explicit examples of such calculations 
are then given in 
Section~\ref{section: examples}.

\section{Calculating the entropy through the canonical ensemble}
\label{section: canonical}

To determine the average thermodynamic energy 
we must calculate  the 
``probabilities,''
$p_j(\nu_j)$,
that were defined in 
Eq.~(\ref{P n1 n12 1}).
The simplest way to do this is through the canonical ensemble,
which  will give us a thermodynamic potential 
that contains all thermodynamic information 
for the subsystem of interest.

A Legendre transform
will then give  the entropy
as a function of the thermodynamic energy.
Because I am interested 
in using a formalism
that is also valid for subsystems
with non-monotonic entropies,
I will use a Massieu function
to derive the entropy\cite{Callen,RHS_book}.
Since a Massieu function might be  somewhat less familiar than 
the more common Helmholtz free energy,
it will be reviewed in the subsection
following that on the canonical ensemble.

\subsection{Canonical ensemble}

If we put the subsystem of interest,
which we will pick to be $j=1$,
into equilibrium with a reservoir that has a known 
continuous density of states,
$\omega_R(E_R)$,
the value of 
$p_1(\nu_1)$
will be determined by the proportionality:
\begin{equation}\label{p1 omega R}
 p_1( \nu_1) 
 \propto
 \omega_R  \left( E_T - E_1 ( \nu_1 )  \right) ,
\end{equation}
where $E_T=E_R+E_1 ( \nu_1 )$
 is the total energy of the reservoir and subsystem~1.
Taking the logarithm of each side and expanding 
$\omega_R$,
gives
\begin{eqnarray}\label{p(nu) 1}
 \ln p_1( \nu_1) 
 &=&
 \ln \omega_R  \left( E_T  \right) 
-
  E_1 ( \nu_1 )
 \left[ \frac{ \partial }{ \partial E_T }
 \ln \omega_R  \left( E_T  \right)   \right]  \nonumber \\
 && - \ln Z' + \dots,
\end{eqnarray}
where $Z'$ is a constant.
The higher-order terms
that are indicated by the dots in 
Eq.~(\ref{p(nu) 1})
are proportional to the 
ratio of $E_1$ to the much larger values of $E_T$,
and in the limit of an infinite reservoir,
they vanish.
Although this calculation is based on finding the mode 
of the probability distribution,
rather than the mean,
the assumption of an infinite reservoir 
is sufficient to make the mean and mode agree.
Interactions between finite subsystems will be discussed elsewhere,
but the effects are proportional to the 
inverse of the number of particles,
and are not measurable for macroscopic subsystems.

Defining 
\begin{equation}
\beta 
=
\frac{ \partial }{ \partial E_T  }
 \ln \omega_R  \left( E_T  \right)   ,
\end{equation}
Eq.~(\ref{p(nu) 1})
 becomes
\begin{equation}
 \ln p_1( \nu_1) 
 =
 \ln \omega_R  \left( E_T  \right) 
-
\beta
  E_1 ( \nu_1 )
 - \ln Z',
\end{equation}
or
\begin{equation}\label{p1 1}
 p_1( \nu_1) 
 =
 \frac{1}{Z'}
\omega_R  \left( E_T  \right) 
\exp \left( -
\beta
  E_1 ( \nu_1 )
  \right)   .
\end{equation}
Since 
$\omega_R  \left( E_T  \right)$
does not depend on $\nu_1$,
we can simplify the expression by defining a 
new constant 
$Z=Z' / \omega_2  \left( E_{1,2}   \right)$,
which gives 
\begin{equation}\label{p1 2}
 p_1( \nu_1) 
 =
 \frac{1}{Z}
\exp \left( -
\beta
  E_1 ( \nu_1 )
  \right)   .
\end{equation}
This is,
of course,
 just the canonical distribution 
for subsystem 1
for an inverse temperature $\beta=1/k_B T$.
The partition function  $Z=Z(\beta)$ in 
Eq.~(\ref{p1 2})
is given by the normalization condition,
\begin{equation}\label{z norm 1}
Z =
 \sum_{\nu_1 }
\exp \left( -
\beta
  E_1 ( \nu_1 )
  \right)           .
\end{equation}
Note that the only property of the reservoir
that enters into this equation is the 
inverse temperature 
$\beta$,
which is the only property of the reservoir 
needed to determine the thermodynamics 
of subsystem~1.

The average energy of subsystem 1,
which can be identified with the 
thermodynamic energy, 
$U_1$,
is
\begin{eqnarray}\label{E1 from p}
U_1 (\beta )
&=&
\sum_{\nu_1}
E_1(\nu_1)
\,
 p_1( \nu_1)       \nonumber \\
 &=&
 \frac{1}{Z(\beta)}
\sum_{\nu_1}
E_1(\nu_1)
\exp \left( -
\beta
  E_1 ( \nu_1 )
  \right)   .
\end{eqnarray}

\subsection{Equilibrium between finite systems}

An important concern for practical applications 
is how 
the values of 
$p_1(\nu_1)$ 
for subsystem 1
in
Eq.(\ref{p1 2})
are affected by replacing the infinite reservoir 
$R$,
used in Eq.~(\ref{p1 omega R}),
by a finite subsystem 2.
The higher-order terms in Eq.~(\ref{p(nu) 1})
no longer vanish,
but they are small.

It is not really necessary for 
Eq.(\ref{p1 2})
to be valid for all values of
$E_1(\nu_!)$.
Because
the values of 
$p_1(\nu_1)$ 
will only 
be significantly different from zero  
within 
 the thermal fluctuations
$\delta E_1$,
it is only necessary for 
Eq.(\ref{p1 2})
to be valid 
over a range of energies of the order of 
$\delta E_1$.
As long as the ratio
$\delta E_1 / U_2$
is small,
Eq.(\ref{p1 2})
will be accurate.
From a consideration of the magnitude of the fluctuations,
the condition of validity 
is that 
\begin{equation}\label{d E1 / U2}
\frac{\delta E_1 }{ U_2}
\approx
 \sqrt{ \frac{  N_1  }{ N_2  (N_1+N_2)}  }
 << 1  .
\end{equation}
This condition will be easily satisfied 
for 
$N_2 >> N_1$.
For the opposite case of 
subsystem 1 being in equilibrium 
with a relatively small
subsystem 2, 
so that 
$N_2 << N_2$,
Eq.~(\ref{d E1 / U2})
becomes
$\delta E_1/U_2 \approx 1/\sqrt{N_2}$.
This condition can  also be satisfied if subsystem 2
is macroscopic.

The conclusion is that it doesn't matter much 
for the distribution of energies 
whether a subsystem  is in equilibrium 
with a large subsystem or a small one --
as long as they are macroscopic.
This is consistent with general experience 
of measured temperatures
and experimental confirmations of the 
zeroth law of thermodynamics.

The next step is to explore the consequences of 
Eqs.~(\ref{p1 2}), (\ref{z norm 1}), and (\ref{E1 from p})
for the entropy
through the use of a Massieu function.

\subsection{Massieu functions}

It is well known that the canonical partition function 
is related to the Helmholtz free energy,
 $F=U-TS$,
by the equation
\begin{equation}
\ln Z ( \beta, V,N ) = - \beta F(T,V,N)  ,
\end{equation}
where $T$ is the temperature,
and
$\beta=1/ k_B T$.
The Helmholtz free energy is,
of course,
the Legendre transform of the 
fundamental relation
$U=U(S,V,N)$
with respect to temperature,
which we will denote as 
$F(T,V,N) = U[T]$,
indicating the Legendre transform 
by the square brackets
around the new variable $T$\cite{Callen,RHS_book}.

Although the use of 
the thermodynamic potentials 
derived from the fundamental equation
$U=U(S,V,N)$
through Legendre transforms
are familiar to all students of thermodynamics,
they are not appropriate for calculating the properties 
of non-monotonic entropy functions.
The reason is that if 
$S=S(U,V,N)$
is not monotonic in $U$,
the function cannot be inverted
to find 
$U=U(S,V,N)$.
However,
we can still use Massieu functions,
which are Legendre transforms 
of
$S=S(U,V,N)$\cite{Callen,RHS_book}.
It will be particularly useful to define a dimensionless entropy,
$\tilde{S}=S/k_B$,
in forming Massieu functions.

From the differential form of the fundamental relation
for $dS$,
we can see that
\begin{equation}
d \tilde{S} = \beta \, dU + \beta P dV - \beta \mu \, dN,
\end{equation}
where 
$P$ is the pressure,  $V$ is the volume,
$\mu$ is the chemical potential, and $N$ is the number of particles.
The inverse temperature $\beta$ is found from the usual equation,
which can be written as
\begin{equation}\label{S[b]=S-bU}
\beta
=
\left(
\frac{ \partial \tilde{S} }{ \partial U }   
\right)_{V,N}  .
\end{equation}

The Legendre transform (Massieu function) is given by
\begin{equation}\label{s[b]= - b F}
\tilde{S}[\beta] = \tilde{S} - \beta U = - \beta \left( U - TS \right) = - \beta F ,
\end{equation}
so that 
\begin{equation}\label{St b = ln Z}
\tilde{S}[\beta] =\ln Z ( \beta, V,N )   .        
\end{equation}
The differential form 
of the Massieu function $\tilde{S}[\beta]$
is then
\begin{equation}
d \tilde{S}[\beta] = - U d \beta + \beta P dV - \beta \mu dN .
\end{equation}
This immediately gives us
\begin{equation}\label{d tilde S / dbeta = - U}
\left(
\frac{ \partial \tilde{S}[ \beta ] }{ \partial \beta  }   
\right)_{V,N}  =
- U   
=
-
\left(
\frac{ \partial  ( \beta F )  }{ \partial \beta  }   
\right)_{V,N} ,
\end{equation}
where the last equality is a well-known thermodynamic identity\cite{Callen,RHS_book}.

\subsection{Inverse Legendre transform of 
$\tilde{S}[\beta]$
to find 
$S(U)$}

To carry out the inverse Legendre transform of 
$\tilde{S}[\beta]$
to find 
$S(U)$,
use 
Eq.~(\ref{d tilde S / dbeta = - U})
to find 
$U=U(\beta)$.
Since $U$ is a monotonic function of 
$\beta$, 
even for a non-monotonic density of states,
we can invert this equation to obtain 
$\beta=\beta(U)$.
From 
Eq.~(\ref{S[b]=S-bU}),
we can find
\begin{equation}\label{St = St b + b U}
\tilde{S} = \tilde{S}[\beta] +\beta(U) U .
\end{equation}
Finally, the entropy with the usual dimensions is given by 
\begin{equation}
S
=
k_B \tilde{S}  .
\end{equation}

\subsection{Alternative calculation of the entropy $S(U)$
from $\beta=\beta(U)$}

Since 
$\beta=\beta(U)$
was found by   inverting
$U=U(\beta)$,
the entropy can also be calculated by integrating 
the thermodynamic identity  in
Eq.~(\ref{S[b]=S-bU})
to find
\begin{equation}
S 
= 
k_B
\int_{U_{min}}^U
\beta(U') dU'
\end{equation}
where 
$U_{min}$
is the minimum value of the thermodynamic energy $U$.
Both methods produce the same results.

The next section discusses
  the application of these methods  
to calculate
$S(U)$ 
for
either a monotonic entropy 
or a non-monotonic entropy.

\section{Examples of the entropy of quantum systems}\label{section: examples}

This section contains explicit calculations of the entropy of quantum systems,
as illustrations of the methods  described in the previous section.
 The first example is a 
system   
  of quantum 
 simple harmonic oscillators,
 which has a monotonic dependence of the entropy
 as a function of energy.    .
 The second example is 
 a system composed of 
two-level objects.
This system 
 has a non-monotonic entropy
and 
illustrates how negative temperatures can 
arise in a quantum system.
For completeness,
I will discuss two kinds of 
two-level systems.

A clear distinction should be made between 
macroscopic systems 
composed of microscopic objects
(simple harmonic oscillators, two-level objects, etc.),
as discussed in this section,
and 
the composite systems 
discussed in Sections
\ref{section: QM}  and \ref{section: canonical},
which were composed of 
macroscopic subsystems.

\subsection{A system composed  of quantum simple harmonic oscillators}

Consider a system composed of  $N$ simple harmonic oscillators.
For simplicity,
we will assume that the frequencies of all oscillators are the same,
so that the Hamiltonian is
\begin{equation}
H_{SHO,k} = \hbar \omega  ( n_k +1/2 )  ,
\end{equation}
where 
$\hbar$ is Planck's constant,
$\omega$ is the angular frequency,
and
$n_k=0,1,2,\dots$
The partition function of the complete system of $N$ oscillators is 
well known to be
\begin{equation}
Z_{\textrm{SHO}}
=
\exp( - \beta \hbar \omega N /2 )
\left( 1 - \exp( -\beta \hbar \omega )  \right)^{-N}   ,
\end{equation}
so that 
\begin{eqnarray}
\tilde{S}_{\textrm{SHO}}[\beta] 
&=&
 \ln Z_{\textrm{SHO}}      \nonumber \\
&=&
-\frac{1}{2} N \beta \hbar \omega - N \ln \left( 1- \exp ( - \beta \hbar \omega ) \right)
\end{eqnarray}
The energy is given by the negative partial derivative of
$\tilde{S}_{\textrm{SHO}}[\beta] $ with respect to $\beta$,
as in 
Eq.~(\ref{d tilde S / dbeta = - U}).
\begin{equation}
U_{\textrm{SHO}}=  
\frac{1}{2}\hbar \omega   N
+
N \hbar \omega  \left(  \exp( \beta \hbar \omega  ) -1 \right)^{-1}
\end{equation}

\subsubsection{Finding $S_{\textrm{SHO}}$ by Legendre transform}

To make the following equations more compact,
define 
\begin{equation}
\hat{U}_{\textrm{SHO}}  = U_{\textrm{SHO}} - \frac{1}{2}\hbar \omega N  .
\end{equation}

The next step is to invert 
$\hat{U}_{\textrm{SHO}}=\hat{U}_{\textrm{SHO}}(\beta)$,
to find 
$\beta=\beta ( \hat{U}_{\textrm{SHO}})$.
To make the notation still more compact,
we will express the results in terms of 
a dimensionless energy variable 
\begin{equation}
x
=
\hat{U}_{\textrm{SHO}}/ N \hbar \omega 
=
\frac{
U_{\textrm{SHO}} -\frac{1}{2} 
\hbar \omega N
}{
N \hbar \omega 
}   ,
\end{equation}
which gives 
\begin{equation}
\beta
=
\frac{1}{\hbar \omega }
\ln  \left[1/x  + 1   \right]
=
\frac{1}{\hbar \omega }
\ln  \left[ \frac{ 1 + x }{x}   \right]
\end{equation}
%


The  inverse Legendre transform,
$\tilde{S}_{\textrm{SHO}}[\beta]$
as a function of 
$x$
is 
\begin{equation}
\tilde{S}_{\textrm{SHO}} [\beta] 
=
-\frac{1}{2} N \ln  \left[1/x  + 1   \right]
+N \ln \left( 1 +x ) \right)     .
\end{equation}

Using 
Eq.~(\ref{St = St b + b U}),
the dimensionless entropy
is found to be
\begin{equation}\label{S SHO 1}
S_{\textrm{SHO}}
=
k_B \tilde{S}_{\textrm{SHO}} 
=
N k_B
\left[
- x \ln x
+ ( 1 + x ) \ln (1+x) 
\right]
\end{equation}
It is readily confirmed that the same result is obtained 
by numerical integration of 
the inverse temperature.

In the limit of $\beta \rightarrow \infty$
(or $T \rightarrow 0$)
the energy goes to its minimum value,
$U_{\textrm{SHO}} \rightarrow \hbar \omega N /2$,
and $x \rightarrow 0$.
In this limit,
$S_{\textrm{SHO}} \rightarrow 0$, as expected
from the third law of thermodynamics
(Nernst theorem).

\subsection{A system of two-level quantum objects}\label{subsection: two level}

The next example illustrates  the 
thermodynamics of a system 
with 
a bounded energy spectrum
and
a non-monotonic entropy.
This system displays negative temperatures.

Consider a collection of $N$
two-level quantum objects,
\begin{equation}
H_{\textrm{2-level}}
=
\epsilon \sum_{k=1}^N   n_k  ,
\end{equation}
where 
$\epsilon$ is the energy difference between the two levels in each object, and
$n_k=0$ or $1$.
The partition function $Z_{\textrm{2-level}}$ can be found by standard methods.
\begin{equation}
Z_{\textrm{2-level}}
= 
\left[ 1 + \exp (- \beta \epsilon )  \right]^N
\end{equation}
This gives
\begin{equation}
\tilde{S}_{\textrm{2-level}}[\beta] 
=
\ln  Z_{\textrm{2-level}}
=
N \ln \left[ 1 + \exp (- \beta \epsilon )  \right]   ,
\end{equation}
and,
using 
Eq.~(\ref{d tilde S / dbeta = - U}),
\begin{equation}
U_{\textrm{2-level}}
=  
N \epsilon \left(  \exp( \beta \epsilon ) +1 \right)^{-1}    .
\end{equation}

\subsubsection{Finding $S_{\textrm{2-level}}(U)$ by Legendre transform}

As before, we simplify  the notation
by defining a dimensionless energy
\begin{equation}\label{y = U 2 level dimensionless}
y = \frac{ U_{\textrm{2-level}} }{ N \epsilon }  .
\end{equation}
The same type of calculation used 
for the simple harmonic oscillators
gives
 the inverse temperature as
\begin{equation}\label{beta 2 level 1}
\beta
=
\frac{1}{\epsilon}
\ln 
\left[ \frac{ 1}{y }  -1   \right]
=
\frac{1}{\epsilon}
\ln  \left[ \frac{ 1-y}{y }   \right]   .
\end{equation}
A little algebra then gives
 \begin{equation}\label{S 2-level 1}
S_{\textrm{2-level}}
=
- N k_B 
\left[
y \ln y
+
(1-y)  \ln ( 1 - y  )
  \right]
 \end{equation}
 As was the case for the system composed of simple harmonic oscillators,
 integration of the inverse temperature produces the same result
 as the Legendre transform.

$S_{\textrm{2-level}}$
 has positive temperatures for $y<0.5$, and 
negative temperatures for $y > 0.5$,
as expected.
The entropy is symmetric 
for
$y \leftrightarrow 1-y$.
The entropy goes to zero in the limits 
$y \rightarrow 0$
and 
$y \rightarrow 1$,
also as expected.

\subsection{Independent Ising spins}

Since a system composed of non-interacting Ising spins
is isomorphic to the two-level system 
discussed above,
 its entropy can be obtained  with little effort.

Consider the Hamiltonian
\begin{equation}
H_{\textrm{spins}}
=
- b \sum_{j=1}^N \sigma_j  ,
\end{equation}
where 
$\sigma_j$ takes on the values $+1$ and $-1$,
and the parameter $b$
represents an applied magnetic field.
The mapping between the variable $n_j$, 
which take on the values 
$0$ and $1$,
and the spins
is 
\begin{equation}
\sigma_j = 2  n_j -1   ,
\end{equation}
and
the energies map as 
\begin{equation}\label{U 2l and spins}
\frac{ U_{\textrm{2-level}}  }{ N \epsilon}
=
\frac{U_{\textrm{spins}} }{2Nb}
+
\frac{ 1 }{2}  .
\end{equation}

Defining  a dimensionless energy 
for the spin system by 
\begin{equation}
r
=
\frac{  U_{\textrm{spins}}  }{ N b }  ,
\end{equation}
Eq.~(\ref{U 2l and spins})
becomes
\begin{equation}\label{y 2L and spins dimensionless}
y
=
\frac{r+1}{2} ,
\end{equation}
where $y$ is the dimensionless energy for the 
two-level system defined in 
Eq.~(\ref{y = U 2 level dimensionless}).
This gives  the entropy of the spin system as
\begin{eqnarray}\label{S spins 2}
S_{\textrm{spins}}
&=&
-
\frac{ N k_B }{ 2 }
\left[
(1-r)  \ln (1-r) \right.     \nonumber \\
&&+
\left.
 (1+r) \ln(1+r)
 -
 2 \ln 2
\right]  .
\end{eqnarray}

$S_{\textrm{spins}}$
 has positive temperatures for $r<0$, and 
negative temperatures for $r>0$,
as expected,
with an obvious symmetry 
for
$r \leftrightarrow -r$.
The entropy goes to zero in the limits 
$r \rightarrow -1$
and 
$r \rightarrow +1$,
also as expected.

A slight modification of the derivations 
given above 
would also allow the calculation of the entropy 
of a system composed of two-level systems
or simple harmonic oscillators 
with arbitrary distributions of energy-level spacings.

Now that these examples of derivations 
of the entropy of quantum systems are complete,
they can be tested against exact results 
and compared with the predictions 
of the volume entropy 
advocated by opponents 
of the concept of negative temperatures.

\section{Comparison of $S_{\textrm{2-level}}$
with the  volume entropy}

Comparisons with the volume entropy as used by its advocates
are somewhat ambiguous 
because they present their ideas 
in terms of a non-analytic expression
for the entropy
of a quantum system\cite{DH_Physica_A_2006,HHD_2014}.    
However, 
their basic claim      
is that the 
logarithm of the
integral of the density of states gives the true entropy,
which always has positive temperatures.
The volume entropy is calculated in the next section 
 using the same procedure
used in 
Ref.~\onlinecite{HHD_2014}.
The predictions of the volume entropy
are then tested against exact results and 
compared with  the expressions of the entropy
derived in Section~\ref{section: examples}.

\subsection{Calculation  
of  the  volume entropy
for a  system composed of two-level objects}

In the  notation
of Ref.~\onlinecite{HHD_2014},
$\tilde{S}_{\textrm{2-level}}$
is related to the density of states
$\omega(y)$ by
\begin{eqnarray}\label{S surface 1}
\tilde{S}_{\textrm{2-level}}
&=&
\ln \omega (y)    \nonumber \\
&=&
- N 
\left[
y \ln y
+
(1-y)  \ln ( 1 - y  )
  \right] ,
\end{eqnarray}
where 
the explicit dependence on the dimensionless energy $y$
is taken from 
Eq.~(\ref{S 2-level 1}).
Since the dimensionless entropy is being expressed 
in terms of a dimensionless energy variable,
$y= U_{\textrm{2-level}}/N \epsilon$,
there is no need for the  energy parameter 
that
Ref.~\onlinecite{HHD_2014}
included in the logarithm.

The dimensionless   volume entropy is given by 
\begin{equation}
\tilde{S}_G (y)
=
\ln \Omega(y)    ,
\end{equation}
where
\begin{eqnarray}\label{Omega 2-level}
\Omega(y)
&=&
\int_0^y 
\omega( y' )  \, dy'         \nonumber \\
&=&
\int_0^y  
(1-y')^{-N(1-y')}
(y')^{-Ny'}   \, dy'     .
\end{eqnarray}

Carrying out the integral in 
Eq.~(\ref{Omega 2-level})
(numerically)
gives
$\tilde{S}_G$.
After this is done,
the fact that
\begin{equation}
\omega (y)
=
\frac{ \partial \Omega (y) }{ \partial y }   
\end{equation}
gives a simple expression for the 
inverse temperature.
\begin{equation}
\beta_G
=
\frac{ \partial \tilde{S}_G (y) }{ \partial y }   
=
\frac{  \omega (y)  }{  \Omega (y) }  
\end{equation}

These explicit expressions for the volume entropy
and its associated temperature 
are tested  against exact results in the next subsection.

\subsection{Comparison between predictions of 
the partition of energy 
between subsystems in equilibrium 
for different entropies}

Consider thermal equilibrium between two two-level subsystems 
with the same level spacing $\epsilon$,
but differing in size by a factor of $f$.
Subsystem 1 contains $fN$ two-level objects,
and subsystem 2 is $1/f$ smaller, with $N$ objects.
Even without calculations,
it is obvious that equipartition of energy 
between the two subsystems in equilibrium
requires 
\begin{equation}\label{U1/fN = U2/N}
\frac{U_1 }{ fN } 
=
\frac{U_2 }{ N }          .
\end{equation}

The prediction of 
$\tilde{S}_{\textrm{2-level}}$
comes from setting the temperatures equal.
Using                       		
Eq.~(\ref{beta 2 level 1}),
\begin{equation}
\beta_j
=
\frac{1}{\epsilon}
\ln 
\left[ \frac{ f_j  N \epsilon }{ U_{\textrm{2-level},j} }  -1   \right]
=
\frac{1}{\epsilon}
\ln 
\left[ \frac{ 1 }{ y_j }  -1   \right]   ,
\end{equation}
for $j=1$ or $j=2$,
and $f_1=f$, while $f_2=1$.
Since 
$\epsilon$ is the same in both subsystems,
it is clear that the condition of equilibrium is 
$y_1=y_2$,
which is equivalent to 
Eq.~(\ref{U1/fN = U2/N}).
This confirms that 
$\tilde{S}_{\textrm{2-level}}$
correctly predicts 
equipartition 
for these subsystems with non-monotonic 
$\omega(y)$.

\begin{figure}[htb]
\begin{center}
 \includegraphics[width=4in]{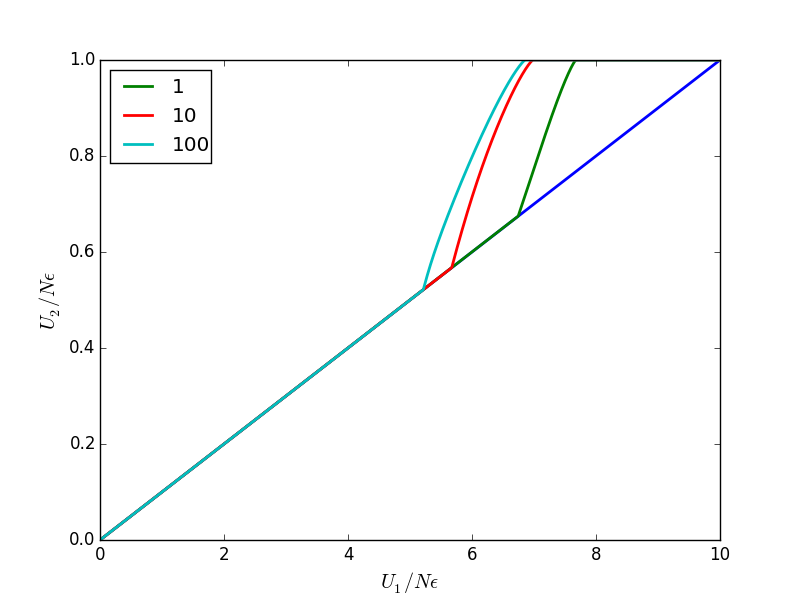}\\
  \caption{Plot of the predictions
  of the  volume entropy
  are compared with those 
  of the entropy given by
   $S_{\textrm{2-level}}$
     in
  Eq.~(\ref{S 2-level 1}) 
  for the distribution of energy 
  between two subsystems of 
  two-level objects
  in thermal equilibrium.    
  The exact condition of 
  equipartition of energy 
  is indicated by the straight line,
  which is also the prediction of 
  $S_{\textrm{2-level}}$.    
  The three curves giving the 
  volume entropy predictions are for subsystem sizes 
   $[N_1,N_2]$ of
     $[10,1]$,      
     $[100,10]$, and       
     $[1000,100]$. 
     In each case, the value of $N_2$ is given in the legend.   
     Larger deviations of the curves for the volume entropy 
     from the exact results
     correspond to larger subsystems.
     Increasing the ratio of the subsystem sizes also 
     increases the violation of equipartition of energy.
  }
  \label{f10_plot}
\end{center}
\end{figure}

The numerical computations for
the equilibrium predictions 
of the volume entropy have been 
have been  carried out 
for test cases 
and plotted 
in 
Fig.~\ref{f10_plot}.
For comparison,
the exact condition
for equilibrium,
which is a straight line
given by
Eq.~(\ref{U1/fN = U2/N}),
is also shown.
This exact equilibrium condition is 
identical to that predicted by
$S_{\textrm{2-level}}$,
as derived in 
Section \ref{subsection: two level}.
For values of the energy
less than half the maximum energy,
all predictions are in good agreement 
with the equipartition of energy.
However,
for values of the energy
more than half the maximum energy,
the predictions of the  volume entropy 
deviate significantly from 
the correct equilibrium conditions.

The origin of the errors made by the volume entropy 
can be traced back to the size dependence of the 
temperature 
$T_G$ 
given by the volume entropy:
the larger the subsystem, the higher $T_G$.
Since subsystem~1 is larger than 
subsystem~2,
it also has a higher value of $T_G$ at equilibrium
than 
subsystem~2.
Equal values of 
$T_G$
in the two subsystems 
can only be achieved if
subsystem~2  has a higher energy 
than appropriate for equilibrium,
leading to the erroneous predictions 
of the volume entropy shown in 
Fig.~\ref{f10_plot}.

\section{Conclusions}

I have demonstrated that the entropy 
of a macroscopic quantum system
is a continuous function 
of the thermodynamic energy,
as opposed to the  step functions and delta functions
proposed by other authors\cite{DH_Physica_A_2006,HHD_2014,HHD_2015,Planck_1901,Planck_book}.

The expressions for the entropy 
of 
simple harmonic oscillators in 
Eq.~(\ref{S SHO 1}) 
and 
two-level systems in
Eqs.~(\ref{S 2-level 1})
and (\ref{S spins 2})
satisfy the postulates of thermodynamics.
They
are fundamental thermodynamic  relations
that completely characterize 
the correct  thermodynamic properties 
of these systems.
Because the fundamental relation 
of a thermodynamic system
can be shown to be unique,
any valid definition 
of the entropy of macroscopic quantum systems
must be equivalent to the one presented in this paper.

An interesting feature of the definition of entropy 
presented here
is that subsystems with incommensurate energy eigenvalues
can be in equilibrium with each other.
It is, of course,
well known experimentally
that differences in microscopic energy-level spacings 
do not prevent equilibration, 
but this fact  has not  been obvious from earlier 
proposed expressions for the entropy of 
macroscopic systems.

Finally,
I have shown that the entropy of  two-level objects 
confirms  the validity of 
the concept of
negative temperatures.

\section*{Acknowledgement}

I would like to thank
Oliver Penrose and Jian-Sheng Wang
for extremely   perceptive and valuable comments.
I  would also like to thank 
Roberta Klatzky 
for many helpful discussions.

\makeatletter
\renewcommand\@biblabel[1]{#1. }
\makeatother

\bibliography{Entropy_citations_2}

\begin{thebibliography}{27}%
\makeatletter
\providecommand \@ifxundefined [1]{%
 \@ifx{#1\undefined}
}%
\providecommand \@ifnum [1]{%
 \ifnum #1\expandafter \@firstoftwo
 \else \expandafter \@secondoftwo
 \fi
}%
\providecommand \@ifx [1]{%
 \ifx #1\expandafter \@firstoftwo
 \else \expandafter \@secondoftwo
 \fi
}%
\providecommand \natexlab [1]{#1}%
\providecommand \enquote  [1]{``#1''}%
\providecommand \bibnamefont  [1]{#1}%
\providecommand \bibfnamefont [1]{#1}%
\providecommand \citenamefont [1]{#1}%
\providecommand \href@noop [0]{\@secondoftwo}%
\providecommand \href [0]{\begingroup \@sanitize@url \@href}%
\providecommand \@href[1]{\@@startlink{#1}\@@href}%
\providecommand \@@href[1]{\endgroup#1\@@endlink}%
\providecommand \@sanitize@url [0]{\catcode `\\12\catcode `\$12\catcode
  `\&12\catcode `\#12\catcode `\^12\catcode `\_12\catcode `\%12\relax}%
\providecommand \@@startlink[1]{}%
\providecommand \@@endlink[0]{}%
\providecommand \url  [0]{\begingroup\@sanitize@url \@url }%
\providecommand \@url [1]{\endgroup\@href {#1}{\urlprefix }}%
\providecommand \urlprefix  [0]{URL }%
\providecommand \Eprint [0]{\href }%
\@ifxundefined \urlstyle {%
  \providecommand \doi  [0]{\begingroup \@sanitize@url \@doi}%
  \providecommand \@doi [1]{\endgroup \@@startlink {\doibase
  #1}doi:\discretionary {}{}{}#1\@@endlink }%
}{%
  \providecommand \doi  [0]{doi:\discretionary{}{}{}\begingroup
  \urlstyle{rm}\Url }%
}%
\providecommand \doibase [0]{http://dx.doi.org/}%
\providecommand \Doi [0]{\begingroup \@sanitize@url \@Doi }%
\providecommand \@Doi  [1]{\endgroup\@@startlink{\doibase#1}\@@Doi}%
\providecommand \@@Doi [1]{#1\@@endlink}%
\providecommand \selectlanguage [0]{\@gobble}%
\providecommand \bibinfo  [0]{\@secondoftwo}%
\providecommand \bibfield  [0]{\@secondoftwo}%
\providecommand \translation [1]{[#1]}%
\providecommand \BibitemOpen [0]{}%
\providecommand \bibitemStop [0]{}%
\providecommand \bibitemNoStop [0]{.\EOS\space}%
\providecommand \EOS [0]{\spacefactor3000\relax}%
\providecommand \BibitemShut  [1]{\csname bibitem#1\endcsname}%
\bibitem [{\citenamefont {Dunkel}\ and\ \citenamefont
  {Hilbert}(2006)}]{DH_Physica_A_2006}%
  \BibitemOpen
  \bibfield  {author} {\bibinfo {author} {\bibfnamefont {J.}~\bibnamefont
  {Dunkel}}\ and\ \bibinfo {author} {\bibfnamefont {S.}~\bibnamefont
  {Hilbert}},\ }\href@noop {} {\bibfield  {journal} {\bibinfo  {journal}
  {Physica A},\ }\textbf {\bibinfo {volume} {370}},\ \bibinfo {pages} {390}
  (\bibinfo {year} {2006})}\BibitemShut {NoStop}%
\bibitem [{\citenamefont {Vilar}\ and\ \citenamefont
  {Rubi}(2014)}]{Vilar_Rubi_2014}%
  \BibitemOpen
  \bibfield  {author} {\bibinfo {author} {\bibfnamefont {J.~M.~G.}\
  \bibnamefont {Vilar}}\ and\ \bibinfo {author} {\bibfnamefont {J.~M.}\
  \bibnamefont {Rubi}},\ }\href@noop {} {\bibfield  {journal} {\bibinfo
  {journal} {J. Chem. Phys.},\ }\textbf {\bibinfo {volume} {140}},\ \bibinfo
  {pages} {201101} (\bibinfo {year} {2014})}\BibitemShut {NoStop}%
\bibitem [{\citenamefont {Sokolov}(2014)}]{Sokolov_2014}%
  \BibitemOpen
  \bibfield  {author} {\bibinfo {author} {\bibfnamefont {I.~M.}\ \bibnamefont
  {Sokolov}},\ }\href@noop {} {\bibfield  {journal} {\bibinfo  {journal}
  {Nature Physics},\ }\textbf {\bibinfo {volume} {10}},\ \bibinfo {pages} {7}
  (\bibinfo {year} {2014})}\BibitemShut {NoStop}%
\bibitem [{\citenamefont {Hilbert}\ \emph {et~al.}(2014)\citenamefont
  {Hilbert}, \citenamefont {H\"anggi},\ and\ \citenamefont
  {Dunkel}}]{HHD_2014}%
  \BibitemOpen
  \bibfield  {author} {\bibinfo {author} {\bibfnamefont {S.}~\bibnamefont
  {Hilbert}}, \bibinfo {author} {\bibfnamefont {P.}~\bibnamefont {H\"anggi}}, \
  and\ \bibinfo {author} {\bibfnamefont {J.}~\bibnamefont {Dunkel}},\
  }\href@noop {} {\bibfield  {journal} {\bibinfo  {journal} {Phys. Rev. E},\
  }\textbf {\bibinfo {volume} {90}},\ \bibinfo {pages} {062116} (\bibinfo
  {year} {2014})}\BibitemShut {NoStop}%
\bibitem [{\citenamefont {Campisi}(2015)}]{Campisi_SHPMP_2015}%
  \BibitemOpen
  \bibfield  {author} {\bibinfo {author} {\bibfnamefont {M.}~\bibnamefont
  {Campisi}},\ }\href@noop {} {\bibfield  {journal} {\bibinfo  {journal} {Phys.
  Rev. E},\ }\textbf {\bibinfo {volume} {91}},\ \bibinfo {pages} {052147}
  (\bibinfo {year} {2015})}\BibitemShut {NoStop}%
\bibitem [{\citenamefont {H\"anggi}\ \emph {et~al.}(2015)\citenamefont
  {H\"anggi}, \citenamefont {Hilbert},\ and\ \citenamefont
  {Dunkel}}]{HHD_2015}%
  \BibitemOpen
  \bibfield  {author} {\bibinfo {author} {\bibfnamefont {P.}~\bibnamefont
  {H\"anggi}}, \bibinfo {author} {\bibfnamefont {S.}~\bibnamefont {Hilbert}}, \
  and\ \bibinfo {author} {\bibfnamefont {J.}~\bibnamefont {Dunkel}},\
  }\href@noop {} {\enquote {\bibinfo {title} {Meaning of temperature in
  different thermostatistical ensembles},}\ } (\bibinfo {year} {2015}),\
  \bibinfo {note} {arxiv.org:1507.05713}\BibitemShut {NoStop}%
\bibitem [{\citenamefont {Purcell}\ and\ \citenamefont
  {Pound}(1951)}]{Purcell_Pound}%
  \BibitemOpen
  \bibfield  {author} {\bibinfo {author} {\bibfnamefont {E.~M.}\ \bibnamefont
  {Purcell}}\ and\ \bibinfo {author} {\bibfnamefont {R.~V.}\ \bibnamefont
  {Pound}},\ }\href@noop {} {\bibfield  {journal} {\bibinfo  {journal} {Phys.
  Rev.},\ }\textbf {\bibinfo {volume} {81}},\ \bibinfo {pages} {279} (\bibinfo
  {year} {1951})}\BibitemShut {NoStop}%
\bibitem [{\citenamefont {Ramsey}(1956)}]{Ramsey_Neg_T}%
  \BibitemOpen
  \bibfield  {author} {\bibinfo {author} {\bibfnamefont {N.~F.}\ \bibnamefont
  {Ramsey}},\ }\Doi {10.1103/PhysRev.103.20} {\bibfield  {journal} {\bibinfo
  {journal} {Phys. Rev.},\ }\textbf {\bibinfo {volume} {103}},\ \bibinfo
  {pages} {20} (\bibinfo {year} {1956})}\BibitemShut {NoStop}%
\bibitem [{\citenamefont {Landsberg}(1959)}]{Landsberg_1959}%
  \BibitemOpen
  \bibfield  {author} {\bibinfo {author} {\bibfnamefont {P.~T.}\ \bibnamefont
  {Landsberg}},\ }\href@noop {} {\bibfield  {journal} {\bibinfo  {journal}
  {Phys Rev.},\ }\textbf {\bibinfo {volume} {115}},\ \bibinfo {pages} {518}
  (\bibinfo {year} {1959})}\BibitemShut {NoStop}%
\bibitem [{\citenamefont {Dunkel}\ and\ \citenamefont
  {Hilbert}(2013)}]{DH_2013}%
  \BibitemOpen
  \bibfield  {author} {\bibinfo {author} {\bibfnamefont {J.}~\bibnamefont
  {Dunkel}}\ and\ \bibinfo {author} {\bibfnamefont {S.}~\bibnamefont
  {Hilbert}},\ }\href@noop {} {\enquote {\bibinfo {title} {Inconsistent
  thermostatistics and negative absolute temperatures},}\ } (\bibinfo {year}
  {2013}),\ \bibinfo {note} {arXiv:1304.2066v1
  [cond-mat.stat-mech]}\BibitemShut {NoStop}%
\bibitem [{\citenamefont {Romero-Rochin}(2013)}]{Romero-Rochin_2013}%
  \BibitemOpen
  \bibfield  {author} {\bibinfo {author} {\bibfnamefont {V.}~\bibnamefont
  {Romero-Rochin}},\ }\href@noop {} {\bibfield  {journal} {\bibinfo  {journal}
  {Phys. Rev. E},\ }\textbf {\bibinfo {volume} {88}},\ \bibinfo {pages}
  {022144} (\bibinfo {year} {2013})}\BibitemShut {NoStop}%
\bibitem [{\citenamefont {Dunkel}\ and\ \citenamefont
  {Hilbert}(2014)}]{DH_NatPhys_2014}%
  \BibitemOpen
  \bibfield  {author} {\bibinfo {author} {\bibfnamefont {J.}~\bibnamefont
  {Dunkel}}\ and\ \bibinfo {author} {\bibfnamefont {S.}~\bibnamefont
  {Hilbert}},\ }\href@noop {} {\bibfield  {journal} {\bibinfo  {journal}
  {Nature Physics},\ }\textbf {\bibinfo {volume} {10}},\ \bibinfo {pages} {67}
  (\bibinfo {year} {2014})}\BibitemShut {NoStop}%
\bibitem [{\citenamefont {U.~Schneider}\ \emph {et~al.}(2014)\citenamefont
  {U.~Schneider}, \citenamefont {Rapp}, \citenamefont {Braun}, \citenamefont
  {Weimer}, \citenamefont {Bloch},\ and\ \citenamefont
  {Rosch}}]{Schneider_et_al}%
  \BibitemOpen
  \bibfield  {author} {\bibinfo {author} {\bibfnamefont {S.~M.}\ \bibnamefont
  {U.~Schneider}}, \bibinfo {author} {\bibfnamefont {A.}~\bibnamefont {Rapp}},
  \bibinfo {author} {\bibfnamefont {S.}~\bibnamefont {Braun}}, \bibinfo
  {author} {\bibfnamefont {H.}~\bibnamefont {Weimer}}, \bibinfo {author}
  {\bibfnamefont {I.}~\bibnamefont {Bloch}}, \ and\ \bibinfo {author}
  {\bibfnamefont {A.}~\bibnamefont {Rosch}},\ }\href@noop {} {\enquote
  {\bibinfo {title} {Comment on `{Consistent} thermostatistics forbids negative
  absolute temperatures'},}\ } (\bibinfo {year} {2014}),\ \bibinfo {note}
  {arXiv:1407.41227v1 [cond-mat.quant-gas]}\BibitemShut {NoStop}%
\bibitem [{\citenamefont {Frenkel}\ and\ \citenamefont
  {Warren}(2015)}]{Frenkel_Warren_2015}%
  \BibitemOpen
  \bibfield  {author} {\bibinfo {author} {\bibfnamefont {D.}~\bibnamefont
  {Frenkel}}\ and\ \bibinfo {author} {\bibfnamefont {P.~B.}\ \bibnamefont
  {Warren}},\ }\href@noop {} {\bibfield  {journal} {\bibinfo  {journal} {Am. J.
  Phys.},\ }\textbf {\bibinfo {volume} {83}},\ \bibinfo {pages} {163} (\bibinfo
  {year} {2015})}\BibitemShut {NoStop}%
\bibitem [{\citenamefont {Dunkel}\ and\ \citenamefont
  {Hilbert}(2104)}]{DH_reply_to_FW}%
  \BibitemOpen
  \bibfield  {author} {\bibinfo {author} {\bibfnamefont {J.}~\bibnamefont
  {Dunkel}}\ and\ \bibinfo {author} {\bibfnamefont {S.}~\bibnamefont
  {Hilbert}},\ }\href@noop {} {\enquote {\bibinfo {title} {Reply to {Frenkel}
  and {Warren} [arxiv:1403.4299v1]},}\ } (\bibinfo {year} {2104}),\ \bibinfo
  {note} {arXiv:1403.6058v1}\BibitemShut {NoStop}%
\bibitem [{\citenamefont {Swendsen}(2002)}]{RHS_1}%
  \BibitemOpen
  \bibfield  {author} {\bibinfo {author} {\bibfnamefont {R.~H.}\ \bibnamefont
  {Swendsen}},\ }\href@noop {} {\bibfield  {journal} {\bibinfo  {journal} {J.
  Stat. Phys.},\ }\textbf {\bibinfo {volume} {107}},\ \bibinfo {pages} {1143}
  (\bibinfo {year} {2002})}\BibitemShut {NoStop}%
\bibitem [{\citenamefont {Swendsen}(2006)}]{RHS_4}%
  \BibitemOpen
  \bibfield  {author} {\bibinfo {author} {\bibfnamefont {R.~H.}\ \bibnamefont
  {Swendsen}},\ }\href@noop {} {\bibfield  {journal} {\bibinfo  {journal} {Am.
  J. Phys.},\ }\textbf {\bibinfo {volume} {74}},\ \bibinfo {pages} {187}
  (\bibinfo {year} {2006})}\BibitemShut {NoStop}%
\bibitem [{\citenamefont {Swendsen}(2008)}]{RHS_5}%
  \BibitemOpen
  \bibfield  {author} {\bibinfo {author} {\bibfnamefont {R.~H.}\ \bibnamefont
  {Swendsen}},\ }\href@noop {} {\bibfield  {journal} {\bibinfo  {journal}
  {Entropy},\ }\textbf {\bibinfo {volume} {10}},\ \bibinfo {pages} {15}
  (\bibinfo {year} {2008})}\BibitemShut {NoStop}%
\bibitem [{\citenamefont {Swendsen}(2011)}]{RHS_8}%
  \BibitemOpen
  \bibfield  {author} {\bibinfo {author} {\bibfnamefont {R.~H.}\ \bibnamefont
  {Swendsen}},\ }\href@noop {} {\bibfield  {journal} {\bibinfo  {journal} {Am.
  J. Phys.},\ }\textbf {\bibinfo {volume} {79}},\ \bibinfo {pages} {342}
  (\bibinfo {year} {2011})}\BibitemShut {NoStop}%
\bibitem [{\citenamefont {Swendsen}(2012){\natexlab{a}}}]{RHS_9}%
  \BibitemOpen
  \bibfield  {author} {\bibinfo {author} {\bibfnamefont {R.~H.}\ \bibnamefont
  {Swendsen}},\ }\href@noop {} {\bibfield  {journal} {\bibinfo  {journal}
  {Found. of Physics},\ }\textbf {\bibinfo {volume} {42}},\ \bibinfo {pages}
  {582} (\bibinfo {year} {2012}{\natexlab{a}})}\BibitemShut {NoStop}%
\bibitem [{\citenamefont {Swendsen}(2014)}]{RHS_unnormalized}%
  \BibitemOpen
  \bibfield  {author} {\bibinfo {author} {\bibfnamefont {R.~H.}\ \bibnamefont
  {Swendsen}},\ }\href@noop {} {\bibfield  {journal} {\bibinfo  {journal} {Am.
  J. Phys.},\ }\textbf {\bibinfo {volume} {82}},\ \bibinfo {pages} {941}
  (\bibinfo {year} {2014})}\BibitemShut {NoStop}%
\bibitem [{\citenamefont {Swendsen}\ and\ \citenamefont
  {Wang}(2015){\natexlab{a}}}]{SW_2015_AJP}%
  \BibitemOpen
  \bibfield  {author} {\bibinfo {author} {\bibfnamefont {R.~H.}\ \bibnamefont
  {Swendsen}}\ and\ \bibinfo {author} {\bibfnamefont {J.-S.}\ \bibnamefont
  {Wang}},\ }\href@noop {} {\enquote {\bibinfo {title} {Negative temperatures
  and the definition of entropy},}\ } (\bibinfo {year} {2015}{\natexlab{a}}),\
  \bibinfo {note} {arXiv:1410.4619 [cond-mat.stat-mech]}\BibitemShut {NoStop}%
\bibitem [{\citenamefont {Swendsen}\ and\ \citenamefont
  {Wang}(2015){\natexlab{b}}}]{SW_2015_PR_E_R}%
  \BibitemOpen
  \bibfield  {author} {\bibinfo {author} {\bibfnamefont {R.~H.}\ \bibnamefont
  {Swendsen}}\ and\ \bibinfo {author} {\bibfnamefont {J.-S.}\ \bibnamefont
  {Wang}},\ }\href@noop {} {\enquote {\bibinfo {title} {The {Gibbs} ``volume''
  entropy is incorrect},}\ } (\bibinfo {year} {2015}{\natexlab{b}}),\ \bibinfo
  {note} {{Phys.} {Rev.} {E}, to be published, arXiv:1506.06911
  [cond-mat.stat-mech]}\BibitemShut {NoStop}%
\bibitem [{\citenamefont {Planck}(1901)}]{Planck_1901}%
  \BibitemOpen
  \bibfield  {author} {\bibinfo {author} {\bibfnamefont {M.}~\bibnamefont
  {Planck}},\ }\href@noop {} {\bibfield  {journal} {\bibinfo  {journal} {Drudes
  Annalen},\ }\textbf {\bibinfo {volume} {553}},\ \bibinfo {pages} {65}
  (\bibinfo {year} {1901})},\ \bibinfo {note} {reprinted in \textit{Ostwalds
  Klassiker der exakten Wissenschaften, Band 206, ``Die Ableitung der
  Strahlungsgesteze''}}\BibitemShut {NoStop}%
\bibitem [{\citenamefont {Planck}(1906)}]{Planck_book}%
  \BibitemOpen
  \bibfield  {author} {\bibinfo {author} {\bibfnamefont {M.}~\bibnamefont
  {Planck}},\ }\href@noop {} {\emph {\bibinfo {title} {Theorie der
  W\"armestrahlung}}}\ (\bibinfo  {publisher} {J. A. Barth},\ \bibinfo
  {address} {Leipzig},\ \bibinfo {year} {1906})\ \bibinfo {note} {translated
  into English by Morton Masius in M. Planck, \textit{The Theory of Heat
  Radiation}, (Dover, New York, 1991)}\BibitemShut {NoStop}%
\bibitem [{\citenamefont {Callen}(1985)}]{Callen}%
  \BibitemOpen
  \bibfield  {author} {\bibinfo {author} {\bibfnamefont {H.~B.}\ \bibnamefont
  {Callen}},\ }\href@noop {} {\emph {\bibinfo {title} {Thermodynamics and an
  Introduction to Thermostatistics}}},\ \bibinfo {edition} {2nd}\ ed.\
  (\bibinfo  {publisher} {Wiley},\ \bibinfo {address} {New York},\ \bibinfo
  {year} {1985})\BibitemShut {NoStop}%
\bibitem [{\citenamefont {Swendsen}(2012){\natexlab{b}}}]{RHS_book}%
  \BibitemOpen
  \bibfield  {author} {\bibinfo {author} {\bibfnamefont {R.~H.}\ \bibnamefont
  {Swendsen}},\ }\href@noop {} {\emph {\bibinfo {title} {An Introduction to
  Statistical Mechanics and Thermodynamics}}}\ (\bibinfo  {publisher}
  {Oxford},\ \bibinfo {address} {London},\ \bibinfo {year} {2012})\BibitemShut
  {NoStop}%
\end{thebibliography}%

\end{document}